# Deep learning-based UAV detection in the low altitude clutter background

Zeyang Wu [1], Wenbo Wang [1*], Yuexing Peng [1]

[1] Department of Information and Communication Engineering, Beijing University of Posts and Telecommunications, Beijing, China, 100876
[*] Corresponding author: wbwang@bupt.edu.cn

**Abstract:** Unmanned aerial vehicles (UAVs) are widely used due to their low cost and versatility, but they also pose security and privacy threats. Therefore, reliable detection for low-altitude UAVs is an important issue. The strong ground clutter makes the radar echoes from small UAVs submerged in noise, resulting in low radar detection reliability. A low-altitude UAV detection method based on deep contrastive learning is proposed to address the above problems: Concretely, a low-altitude UAV radar echo model under low-altitude clutter interference is first established. Based on the echo components and the UAV Doppler domain identifiable mechanism, a time-frequency transformation method combining ZAM transform and morphological operations is used to suppress the ambiguity problem under clutter. Then feature extraction and fusion method introducing contrast learning is utilized to suppress non-target ground clutter interference. Finally, a detector relying on semantic features is designed for the reliable identification of low-altitude UAVs. The experiments carried out on both real and simulated data confirm that the proposed method can effectively suppress ground clutter and reliably extract recognizable semantic features of UAVs. The proposed method achieves lower false and missing alarms compared with recent state-of-art solutions and improves the detection accuracy by more than 5% for the same signal-to-noise ratio, which effectively improves the detection reliability.

## 1. Introduction

Recently, small UAVs have been increasingly demanded in numerous applications such as communications, aerial photography, and remote sensing [1]. While UAVs are convenient for outdoor missions, they also raised serious security and privacy leakage threats for certain aerial photography with commercial secrets and sensitive areas, as well as illegal flying in flight safety-threatening areas such as airports. Therefore, robust and adaptive detection of low-observable UAVs has important research value and practical significance.

A group of existing UAV detection methods is generally conducted with different technologies including radar, radio [2], and photos [3]. Radar detection methods are highly accurate, all-weather, and sound reliable, thus becoming the dominant technology for UAV detection. Current UAV radar-based detection techniques mainly include passive and active radar detection strategies. Passive radar-based detection methods use external radiation sources [4] (digital TV signals, synchronous satellite communication signals, etc.) or signals emitted by the target itself to achieve target detection. Due to the relatively low power of the external radiation source signal, the received signal containing the target characteristics is weak in complicated electromagnetic environments, resulting in low detection performance. Active radars actively transmit electromagnetic signals and receive reflected echoes from the target and have anti-interference capability. [5] obtains radar cross-section (RCS) of UAV and calculates 2D, 3D ISAR images to detect UAV. [6] captures suspicious UAV targets through GMM-Iteration and the track-before-detect (TBD) algorithm to detect UAVs. However, the above methods based on specific parameters require a priori knowledge of UAV characteristics, and the lack of uniform standards for actual small and medium-sized UAVs will result in inflexible detection models. Another representative group of moving target detection methods is generally based on the constant false alarm rate (CFAR) detector and its improved variants [7]. However, Phase-Coherent Integration methods based on pulse radar generates distance unit jitter and Doppler frequency unit migration with long accumulation times. What's more, CFAR tends to produce the masking effect [8] in multitarget detection scenes resulting from the large threshold caused by strong targets.

Time-frequency distribution (WVD, STFT, FRFT) [9] is a powerful tool for processing nonstationary signals, and detecting performance can be improved by converting and reconstructing a one-dimensional signal into a two-dimensional feature. Extracting the time-frequency features from radar echo signals provides unique doppler motion characteristics of the UAV. However, there are still some shortcomings in the time-frequency detection scheme: the time-frequency transform crossing terms lead to blurred time-frequency distributions, and the resolution increment in both time and frequency domain is a contradictory issue, which leads to poor recognition accuracy [10]. Although signal decomposition (e.g., EMD, wavelet decomposition) can suppress the cross term by splitting the full frequency band into sub-bands [11], the quality of the time-frequency spectrogram is still degraded due to the low-altitude multi-path effect of the radar echoes and clutter interference.

The above-mentioned traditional radar-based detection solutions rely on manual feature extraction, which requires a high level of prior knowledge on behalf of the researcher. Meanwhile, conventional single features cannot adequately represent complex signals. Currently, deep learning models becomes popular due to their excellent end-to-end learning capabilities. Knowledge-driven neural networks reduce the human cost of feature extraction and address the lack of flexibility of traditional methods because of model parameter fixation. Taking advantage of its capabilities in image recognition and object detection [12], [13], DL has been preliminarily explored in radar target recognition and detection. [14] use SVM to achieve great detection performance in small data training mode, however, it is difficult to train for feature distribution with missing features and non-linearity in large data volume(sets). [15] uses CA-CFAR to initially detect clutter traces and classifies them through a recurrent convolutional neural network, and removes the clutter traces to obtain the UAV detection results. However, the need for large data volumes during training and the lack of decoupling of different features of the echo signal in a single feature mapping model limit the performance of deep learning-based methods at low signal-to-noise ratios.

In general, the small size, strong maneuverability, and weak radar scattering cross-sectional area (RCS) of UAVs increase the difficulty of recognition. Meanwhile, the ground clutter interference at low altitude causes drastic changes in radar echo amplitude and phase, resulting in a large number of false alarms and missed detections of UAV. In this article, a low-altitude fixed-wing UAV detection scheme based on deep contrast learning and time-frequency image recognition is proposed. Concretely, the contributions of this paper are threefold.

1. An impulse Doppler radar echo model of UAV in the ground clutter background is derived based on the noise of the low-altitude ground clutter channel and the multi-path effect under specular reflection. By analysing the echo components, a kernel function-based ZAM transform [16] is used to obtain the radar echo time-frequency image of the moving UAV. An image-based morphological open operation pre-processing is developed to alleviate the time-frequency ambiguity and Doppler frequency migration caused by amplitude-phase interference of multi-path effect and velocity change.
2. A novel feature extraction and fusion strategy combining contrast learning and autoencoder is designed to remove non-target radial echo interference in the time-frequency domain while achieving feature dimensionality reduction.
3. A gated recurrent recursive unit (GRU) based network is utilized to identify extracted UAV motion feature representation. The method achieves promising UAV detection accuracy at low signal-to-noise in the background of ground clutter, and the false alarm and missed detection rates are reduced at the same time.

The rest of the paper is organized as follows: Section 2 provides a low-altitude multi-path-based UAV echo model for linear modulated pulse radar and a time-frequency image acquisition method for UAV detection tasks based on ZAM transformation and image pre-processing. Section 3 describes the UAV time-frequency domain feature extraction and recognition method and presents loss functions, training techniques, and data processing procedures for the entire detection network. Construction of the experiment and the performance of the proposed method verified on real and simulated datasets is shown in Section 4. Results and conclusions are drawn in Section 5.

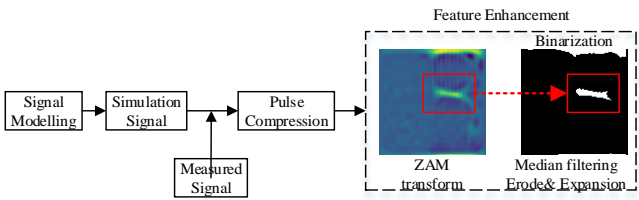

**Fig. 1.** *UAV radar echo signal modeling and time-frequency domain feature transformation process*

## 2. Materials and Methods

For a clear and complete description of the proposed method, this section provides a low-altitude multi-path-based UAV echo model for linear frequency modulated (LFM) pulse radar. How to acquire the time-frequency image based on ZAM transform and image pre-processing from the acoustic data is also presented. As can be seen from Fig. 1, The dynamic radar signal echo model of low-altitude UAV intrusion underground clutter interference is established according to the single-station pulsed Doppler radar regime and multi-path electromagnetic transmission theory [17]. The simulation data constructed in this study is consistent with the scenario of real measured data, and the target Doppler features are extracted by analysing the identifiable mechanism of the target Doppler features in the echoes and transforming the received signal to the time-frequency domain using the ZAM transform.

### 2.1. Scene analysis and signal modeling

LFM is a common modulation signal method for pulsed radar. Assume that the moment of emission of the $m$th pulse is $t_m = mT$ and T denotes the pulse repetition interval (PRI), the mathematical transmitted pulse signal expression is:

$$S(t) = A(t - mT)\, exp[j(2\pi f_t t + \phi_0)] \quad (1)$$

, where $A(t)$ is the pulse envelope, $f_t = f_c + K\frac{t}{2}$, $K = B/T$ is the slope of the linear FM signal, and $f_c$ is the carrier frequency.

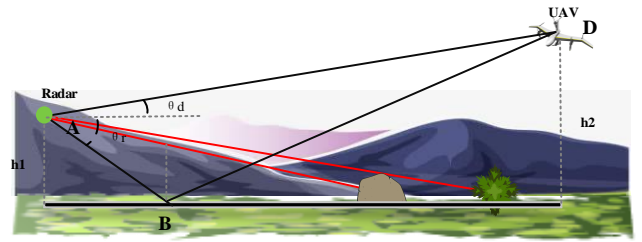

**Fig. 2.** *Multipath propagation model for low-altitude UAV radar signals*

According to electromagnetic transmission theory, the multi-path effect [18] in single station radar causes echoes interference during low angle detection, causing the amplitude and phase of the combined signal to decay. As shown in Fig.2, the transmit signal and the return signal follow two possible paths, resulting in four possible signal returns from transmitting to receive. denotes the distance between the target and the radar, and the target direct path return signal is as follows:

$$S_d(t) = kA(t - mT - \frac{2R(mT)}{c})\, exp[j(2\pi f_t(t - \frac{2R(t)}{c}) + \phi_0)] \quad (2)$$

, assume that at the moment $t = 0$, the target is traveling forward with a speed of $v$m/s at an angle $\phi$ to the radar line of sight, then take the distance $R(t)$ between UAV and Radar for analysis, that is:

$$R(t) = \sqrt{R_0^2 + (v\cos\phi\, t)^2} \quad (3)$$

According to the Rayleigh criterion [19], the multipath signal at this time is mainly a specular reflection signal, mainly concentrated in the vicinity of the geometric reflection point. Assuming that the horizontal distance between the target and the radar is at a certain moment, the geometric relationship in the diagram is used to calculate the difference in distance between the two paths and the expression for the combined signal.

The difference in distance between the two propagation paths:

$$l_d = \sqrt{h^2 + (h_1 + h_2)^2} - R \quad (4)$$

Echo signals at the receiver:

$$S_{d\_multipath}(t) = S_d(t) + S_{d_1}(t) + S_{d_2}(t) + S_{d_3}(t)$$
$$= kA(t - mT - \frac{2R(mT)}{c})\exp[j(2\pi f_t(t - \frac{2R(t)}{c}) + \phi_0)] +$$
$$2\rho kA(t - mT - \frac{2R(mT)}{c} - t_\Delta)\exp[j(2\pi f_t(t - \frac{2R(t)}{c} - t_\Delta) + \phi_0 + \phi_\Delta + \phi_\rho)] +$$
$$\rho^2 kA(t - mT - \frac{2R(mT)}{c} - 2t_\Delta)\exp[j(2\pi f_t(t - \frac{2R(t)}{c} - 2t_\Delta) + \phi_0 + 2\phi_\Delta + 2\phi_\rho)] \quad (5)$$

If $A(t)$ is a rectangular signal of amplitude $A$, for simplicity of calculation, consider only the real part of the signal in the following cases:

$$real(S_{d_{multipath}}(t)) \approx kA\cos(\alpha) + 2\rho kA\cos(\alpha + \beta) + \rho^2 kA\cos(\alpha + 2\beta) = kAB\cos(\alpha + \phi_B) \quad (6)$$

, where $\alpha = 2\pi f_t(t - \frac{2R(t)}{c}) + \phi_0$, k is determined by the electromagnetic reflection coefficient of the material coating of the UAV body, $B = (1 + \rho\cos(\beta))^2 + \rho^2\sin^2(\beta)$ is the change in signal amplitude due to specular multipath interference, $\phi_B$ is the phase change due to interference for multi-path effects, $t_\Delta = l_d/c$ and $\phi_\Delta$ are the phase and time differences due to the difference in distance between the reflective and direct paths, respectively. Reflected signals are assumed to undergo a change in amplitude of $\rho$ and a change in phase of $\phi_\rho$. $\beta = \phi_\Delta + \phi_\rho - 2\pi f_t t_\Delta$ varies with the position of the moving UAV target and $\rho$ and $\beta$ jointly determines the trend of variable $B$.

The intensity of ground clutter is controlled by the magnitude of the reflection coefficient $\rho$, which characterizes the roughness of the reflecting surface, according to the Fresnel model [20]:

$$\rho = \rho_0 \rho_s$$
$$\rho_s = \begin{cases} \exp[-2(2\pi\Gamma)^2], 0 \leq \Gamma \leq 0.1 \\ \frac{0.812537}{1 + 2(2\pi\Gamma^2)}, \Gamma \geq 0.1 \end{cases} \quad (7)$$

, where $\rho_0$ is the Fresnel reflection coefficient for an ideal plane with the same medium as the ground, $\rho_s$ is the specular reflection correction factor. Fig.3(a) shows the relationship between B and β for different choices of ground reflection coefficients when β varies uniformly on 0-2Π. However, the actual multipath interference often satisfies the Riley distribution, when β is Rayleigh distributed over 0-2Π with time, B varies as in Fig. 3(b). Through analysis, it can be obtained that the amplitude of the UAV echo signal varies drastically with the apparent distance between the target and the radar due to the interference of the multi-path effect, and no longer satisfies the one-to-one arrival time-distance mapping relationship.

At the same time, in addition to the UAV target echoes shown in the black line in Fig. 2, radar echoes from some non-

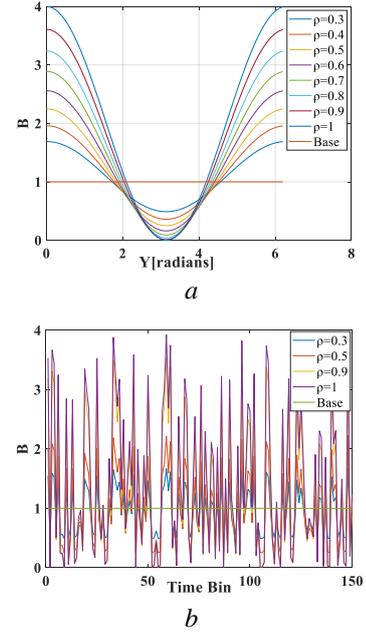

*a*

*b*

***Fig. 3.*** *Plot of B versus β when A takes different values*

target objects on the ground and radial echoes from the ground during radar low pitch tracking are sometimes scattered into the radar receiver system to form ground clutter interference [21]. The red line in Fig. 2 indicates the surface non-target echo, and the combined signal received by the radar receiver at this time is expressed as：

$$S_r = S_{d_{multipath}}(t) + S_u(t) + S_n(t) \quad (8)$$

, where $S_u(t)$ is the non-target echo from the ground surface, and $S_n(t)$ is the Gaussian white noise. The above analysis reveals that the multi-path effect of radar propagation and the interference from ground non-target echoes severely degrade the echo signal-to-noise ratio and affect the detection performance. In this paper, a feature extraction network is designed to suppress the amplitude fading and the interference of non-target echo components under the multi-path effect respectively.

### 2.2. Signal Preprocessing and Time-frequency transform

Since the captured radar echo signal of a UAV with a low signal-to-noise ratio contains a large number of components, processing it directly will increase unnecessary calculations. As shown in Fig. 1, the three components of preprocessing are described as follows:

1. *Pulse Compression*: When using specially modulated waveforms such as linear frequency modulation (LFM) and meeting the conditions of a large time-bandwidth product, matched filtering is equivalent to pulse compression. Pulse compression means that the radar uses a wide pulse signal when transmitting and outputs a narrow pulse after receiving and processing the return signal. Both the range resolution of radar and SNR can be enhanced through pulse compression. The time-domain filter for pulse compression is:

$$h(t) = MS^*(t_0 - t) \quad (9)$$

, where $M$ is the gain, $t_0$ is the time delay of the physical device, and $S^*(t)$ is the conjugate transpose of the transmitted LFM signal $S(t)$. Suppose that the digital matrix filter after sampling and rearrangement is $H_d$, and then, the outputs of the matched filter are:
$$S_d = S_{echo} * H_d \quad (10)$$
, where represents the convolution operation.

2. *Time-frequency transform*: It is known from equation 5 that for UAV targets, their different radial velocities relative to the radar produce different doppler frequencies. Therefore, time-frequency transform specifically utilizing ZAM transform is applied to transfer the pulse-compressed echo signal to the time-frequency domain to obtain the Doppler information of the UAV target and enhance the signal processing gain. The ZAM transform is defined as [22]:
$$ZAM_s(t,f) = \int \phi(\theta,\tau) \int_{t-\frac{|\tau|}{2}}^{t+\frac{|\tau|}{2}} s\left(u+\frac{\tau}{2}\right) s^*\left(u-\frac{\tau}{2}\right) e^{-j2\pi f\tau} du d\tau \quad (11)$$
, where $s^*(t)$ is the conjugate transpose of $s(t)$ and $\phi$ is the kernel function which is expressed as $\phi(\xi,\tau) = g(\tau)|\tau|\frac{sin(a\xi\tau)}{a\xi\tau}$.

The ZAM transform is a time-frequency transform that is less affected by cross-talk terms and can balance the time-frequency resolution of the image. The autocorrelation operation can suppress non-coherent noise in low signal-to-noise scenarios and amplify signal features with periodic components to obtain better time-frequency feature aggregation. Fig. 4 shows the ZAM time-frequency images of radar echoes for two UAV targets intrusion at a signal-to-noise ratio of 3dB for the multipath effect analysed in section 2.1.

As can be seen from the figure, the time-frequency image of the UAV radar echo contains both a ground clutter component unrelated to the target and a target Doppler component, which is consistent with the mechanism deduced from equations 1-8. Meanwhile, the multi-path effect makes the signal amplitude decay, resulting in the ambiguity of the target Doppler features in the time-frequency images, and the fuzzy function of the ZAM transform leads to the interference of cross-terms in the time-frequency image.

3. *Median filtering, Binarization, and Morphological processing*: To improve the resolution of the target Doppler frequency lines in the image, this paper introduces median filtering and Binarization pre-processing [23] to initially suppress the cross-terms generated by the target signal and low-frequency clutter while using morphological open operations [24] to reduce the degree of Doppler ambiguity. The morphological open operation is particularly suitable for dealing with the shape of image elements. It simplifies and maintains the main shape features of the image, and the morphological open operation performed on a binary image is represented as follows:
$$\boldsymbol{M = (B(t,f)\ominus A_1) \oplus A_2} \quad (\boldsymbol{12})$$
, where $B(t,f)$ denotes the binarised time-frequency image, A1 and A2 denote the corrupted and inflated structuring units respectively, $\Theta$ denotes the erode operation, and $\oplus$ denotes the expansion operation. The expand operation causes an expansion of the highlighted areas in the image; the erode operation causes a contraction of the highlighted areas in the image.

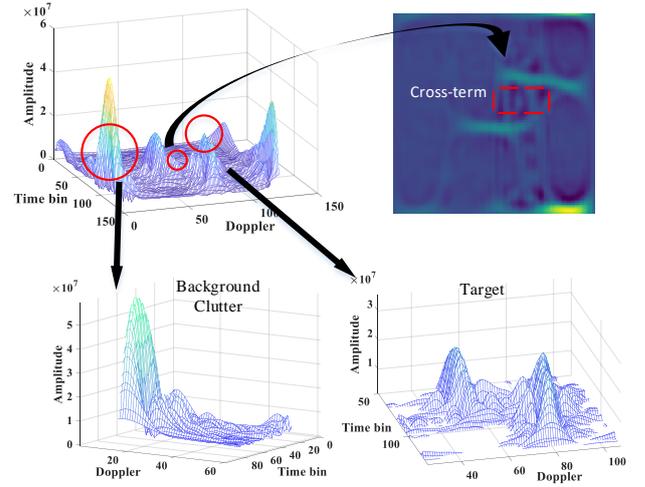

*Fig. 4. Time-frequency images of radar echo signals for background clutter, and cross-term when SNR=3dB*

## 3. Feature Extraction

As Fig.4 shows, The UAV radar echo signal has a ground clutter feature - Z and a target feature - S. The clutter feature refers to the received non-target echoes, the radial echoes reflected directly from the ground during low pitch tracking and white noise, while the target feature refers to the feature containing the UAV motion information. The feature vector p(z,s) obtained by directly downscaling the time-frequency image has "feature entanglement" [25] in the feature space - the ground clutter features and the target Doppler features needed for detection are both included in the downscaled feature vector, which adds additional recognition difficulties. This results in a high false alarm and missed detection rate, as low-frequency clutter spectral lines may be identified as UAV targets. Therefore, a combination of contrast learning and self-encoder-based methods is utilized for feature extraction.

### 3.1. Contrastive learning and Autoencoder

Accurately characterizing data from high-dimensional image signals is extremely difficult, and an effective approach is to replace the predicted loss at each pixel point in the input space with a distance loss in the low-dimensional feature space. A contrast learning framework was generated in response to the inspiration of this idea. Contrastive learning is initially a feature characterization method based on semi-supervised contrastive objective constraint functions [26] and has achieved excellent performance in image and video processing in recent years. The contrastive loss is defined as:
$$L_i^{self} = -log \frac{exp(sim(z_i \cdot z_{j(i)})/\tau)}{\sum_{k=1}^{2N} 1_{i\neq k} exp(sim(z_i \cdot z_k)/\tau)} \quad (13)$$
, where $sim(z_i \cdot z_{j(i)})$ denotes the cosine distance between $Z_i$ and $Z_j$, $\tau$ indicates temperature parameter.

Recently, the fusion of contrast learning with supervised mechanisms for scenarios has been shown to better characterize the feature space of different classes of data than constrained losses such as cross-entropy [27].

The autoencoder network maps an augmented image x to a representation vector $r = e(x) \in R$ which is an asymmetric neural network that learns data features by minimizing reconstruction errors [28]. The target output value of the autoencoder output layer is the input value of its input layer, thus no labelling information of the data is required and it is an unsupervised learning method.

Fig. 5 shows the framework of the proposed method. The feature extraction network consists of two parts: A time-frequency image of radar echoes with unmanned aircraft and containing only non-target ground clutter is fed into a supervised comparative learning network to obtain a representation vector under the constraints of two types of labels: target and non-target. This vector is then combined with the features extracted by the self-encoder for feature recognition. The ground clutter features, which are not useful for UAV recognition, are eliminated by contrast loss, and the reconstruction loss of the self-encoder is used to achieve feature dimensionality reduction at the pixel level and to suppress the time-frequency ambiguity caused by signal fading under multi-path interference. At the same time, this network structure compensates for the lack of ability of unsupervised learning networks to decouple multi-dimensional features. Sections 3.1- (1) and 3.1- (2) present the details of the two modules.

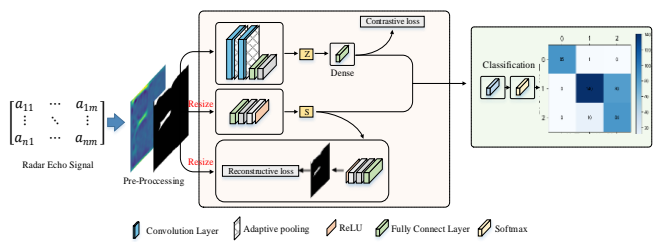

**Fig. 5.** *Complete target detection framework of our method*

1. *Encoder relying on supervised contrastive learning*

The time-frequency images of radar signals with UAV target echoes contain target features as well as clutter features, while the time-frequency images of radar signals without targets contain only low-frequency clutter features, so ground clutter feature cancellation is achieved using a contrast loss-based convolutional neural network with or without targets as an important characterization feature. As shown in Fig. 6 is the structure of a contrastive learning-based encoder consists of three parts: The first part is the data input module: The minibatch was generated by randomly selecting the data. 2N randomly selected data after the data preprocessing in Section 1 formed N sets of time-frequency image pairs [29], $\{x_k, y_k\}$ k = 1... n, where $x_k$ (k = 1... n) are data with UAV targets in the same scene in the dataset, $y_k$ (k = 1... n) is the data with non-target ground clutter only. Where the data with targets in the simulated and measured datasets contain single or dual-target scenarios, both of which are labeled 1, and the rest of the data without targets are labeled 0. Afterward, two encoders are utilized for embedding the time-frequency images into representing vectors. Finally, the normalized representation vectors Z and Z' of the encoder output in 2 are mapped using the multilayer perceptron (MLP) [30] to a vector $z = p(r) \in R$ used to compute the contrast loss, where R is the unit hypersphere. the size of the MLP hidden layer is 2048 and the size of the output vector is 128.

2. *Autoencoder based on image reconstruction*

The original time-frequency image is fed into this encoder after a morphological open operation (swelling erosion) and the reconstructed image is output. The encoder learns the "target features" of the drone echoes in the feature space through reconstruction losses, suppressing frequency cross-terms and image noise.

The configuration details of the whole feature extraction network are listed in Table 1. After combining the feature vectors extracted by the two encoders, the final feature space representation form of the time-frequency image signal is output through a Dense layer, completing the data dimensional compression and feature extraction. Our method encourages the encoder to give closely aligned representations to all entries from the same class, resulting in a more robust clustering of the representation space and this will be proofed by our experiments in Section 4.

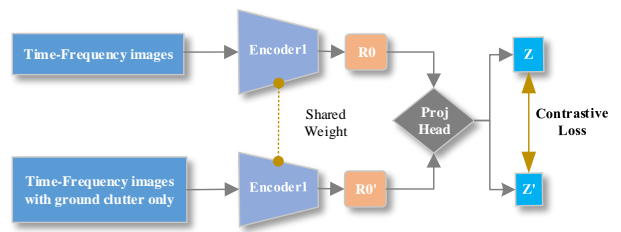

**Fig. 6.** *Contrastive learning-based clutter cancellation network*

### 3.2. Detection network based on recurrent neural network

The length of time a drone is exposed to radar varies with its speed and the scanning speed and viewpoint of the radar, so the length of the echo pulse sequence varies as well. Gated recurrent recursive unit (GRU) networks [31], which take a recursive form in time, pass through the same neuron model for different moments of input, so that the model itself is not specific to a particular length of the sequence while being insensitive to where data features appear in the time dimension. This adaptability makes it flexible enough to handle drone echoes of different lengths without the need to train a new model for each scene separately, making it easier to train and less voluminous than RNN & LSTM networks [32].

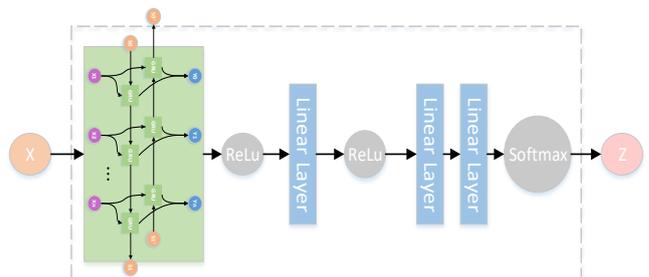

**Fig. 7.** *Architecture of detection network based on GRU*

As shown in Fig. 7, this paper uses a gated recurrent unit (GRU)-based network to classify the feature vector after dimensionality reduction and feature extraction of the self-encoder data to distinguish whether there is an intrusion of a UAV target and, in the presence of a target, whether it is a single target or two targets. As the radar echo signal is complex, the system is an orthogonal two-channel system with the time-frequency images of the real and imaginary signals as the two dimensions of the GRU input. The hidden layer contains 100 bi-directional long and short-term memory units, and the fully connected layer and the output size of the SoftMax layer [33] is 2.

**Table 1** Comparison among encoders and decoders

|  | Layer | Kernel Size | Kernel number | Stride | Padding | Input size | Output size |
|---|---|---|---|---|---|---|---|
| **Encoder1** | Conv1 | 3 x 3 | 8 | 2 | 0 | 128 x 128 | 63 x 63 x 8 |
|  | Adap pool1 | 3 x 3 | - | 1 | 0 | - | 21 x 21 x 8 |
|  | Conv2 | 3 x 3 | 16 | 2 | 0 | 21 x 21 | 10 x 10 x 16 |
|  | Adap pool2 | 3 x 3 | - | 1 | 0 | - | 3 x 3 x 16 |
|  | FC1 | - | - | - | - | 144 | 50 |
|  | FC2 | - | - | - | - | 50 | 20 |
| **Encoder2** | FC3 | - | - | - | - | 28 x 28 | 512 |
|  | FC4 | - | - | - | - | 512 | 128 |
|  | FC5 | - | - | - | - | 128 | 30 |
|  | ReLU | - | - | - | - | 30 | 30 |
| **Decoder** | FC6 | - | - | - | - | 30 | 128 |
|  | FC7 | - | - | - | - | 128 | 512 |
|  | FC8 | - | - | - | - | 512 | 28 x 28 |
|  | ReLU | - | - | - | - | 28 x 28 | 28 x 28 |

### 3.3. Loss functions and training details

Similar to most object detection networks based on images, the loss function of our detection network is also a multitask loss. Let i, j be the index of representing vector of time-frequency images from contrastive encoder mentioned in Section-A-(1), $z_i$ and $z_j$ are the representing vectors of time-frequency images in the minibatch that have the same label $\tilde{l}_i$, as the index i, respectively while $z_i$ and $z_k$ have the different label. Then the contrastive loss function for the encoder in Section-3.1-(1) is:

$$L_{total}^{sup} \sum_{i=1}^{2N} L_i^{sup} \quad (14)$$

$$L_i^{sup} = \frac{-1}{2N_{\tilde{l}_i} - 1} \sum_{j=1}^{2N} 1_{i \neq j} \cdot 1_{\tilde{l}_i = \tilde{l}_j} \cdot \log \frac{exp(sim(z_i \cdot z_j)/\tau)}{\sum_{k=1}^{2N} 1_{i \neq k} \cdot exp(sim(z_i \cdot z_k)/\tau)}$$

, where $N_{\tilde{l}_i}$ is the total number of images in the minibatch that have the same label $\tilde{l}_i$. $\tau > 0$ is a scalar temperature parameter.

The reconstruction loss of the autoencoder in Section-3.1-(2) based on the reconstruction constraint of the time-frequency images is adopted:

$$L_r = E_{E_i \in E} \left[ \sum_{i=1}^{m} \left\| E_i - \hat{E}_i \right\|_2 \right] \quad (15)$$

, where $E_i$ and $\hat{E}_i$ are the original time-frequency images input to the encoder and the time-frequency images reconstructed by the decoder, respectively.

Based on the cross-entropy in information theory, the classification loss for feature detection networks based on GRU network is the logarithmic loss as:

$$L_c = -\frac{1}{n} \sum_{k=1}^{n} q(i) \log(p(i)) \quad (16)$$

, where n is the number of images involved in one iteration of training and $p(i) \in [0, 1]$ be the predicted probability of the category to which the input data belongs. $q(i)$ is the corresponding ground-truth label.

The total number of epochs trained reached 8192, and the model optimizer used Adam with a learning rate of 0.0004. It was observed experimentally that for a fixed epoch, the model converged more effectively when training supervised contrast loss using a larger learning rate and a smaller batch size. To address the problem of inconsistent convergence rates of contrast loss and reconstruction loss, the process of training the self-encoder takes two encoders to be trained in k alternations to ensure that the loss of the whole model converges more fully and uniformly during the gradient descent process. The implementation is carried out with PyTorch.

## 4. Performance Analysis, Results and Discussions

### 4.1. Data Set Description

To evaluate the UAV detection performance of the proposed method, two groups of data are used in the experiments. The first is real measured data jointly published by two scientific institutions [34]: It is a monostatic radar detection signal from a low-altitude UAV obtained in an outdoor scenario. The dataset covers data segments with different signal-to-noise ratios, single UAV targets, and two targets, each segment containing a radar pulse sequence of a certain duration, as well as the corresponding distance wave gate file and annotated truth file. Fifteen segments were selected to validate the performance of the proposed model. 64% and 16% of the data in each data segment are used for independent training and testing of the models respectively, while 20% of the data are used for inter-model performance comparison. The details are shown in Table 2.

**Table 2** Experiment data description

| Basic information | Parameters | Segments | Scene |
|---|---|---|---|
| Carrier Frequency | 35GHz | 1/4/7/10 | SNR=10-20dB, targets from far to near or near to far |
| Modulation | LFM | 2/5/8/11 | SNR=7-10dB, target from far to near or near to far |
| Pulse repetition frequency (PRF)/KHz | 32 | 3/6/9/12 | SNR=3-7dB, targets from far to near or near to far |
| Noise | 3-20dB | - | mean=0, and the standard deviation depends on SNR |

The second data is a MATLAB simulation dataset based on the multipath effect and ground clutter interference of low altitude UAV signal propagation introduced in Section II, where the scenario settings are consistent with the real data. The radar is a monostatic radar system with a height of about 40–50 m and the radial velocity of the UAV varieties between 10m/s-45m/s. The pulse repetition frequency (PRF) is set to 2KHz, the carrier frequency is 35 GHz, and the carrier signal is an LFM signal with a pulse width of 50 μs and a bandwidth of 10 MHz Trees and low walls are set on the ground within radar scanning range and the radar echoes from these targets constitute non-target ground clutter interference.

### 4.2. Comparison Methods

Two different UAV detection methods are used for comparison in the experiments. The first method is a principal component analysis (PCA) and support vector machine (SVM)-based classifier for UAV detection proposed in [35]. PCA is a dimensionality reduction technique, which is the least mean square linear projection of the data in the main subspace, which is a low-dimensional linear subspace. The dimensioned data is fed into an SVM-based classifier to solve for the decision surface that correctly classifies the sample with the largest interval. The second method is CA-CFAR proposed in [36], which is a conventional method following pulse phase reference accumulation and dynamic calculation of judgment thresholds. The last one is a DL-based method proposed by [37], which uses the CNNs composed of two convolutional layers, two pooling layers, and two FC layers with only a classification head for radar UAV detection.

### 4.3. Evaluation metrics

Different evaluation criteria are utilized to compare the performance of the comparison method and the proposed method in this paper in terms of feature extraction and detection accuracy of radar signals, respectively.

As for feature extraction, on the one hand, we plotted the Range-Doppler (RD) maps for the CA-CFAR method and the time-frequency images for short-time Fourier transform (STFT) algorithm [38] used in [37] and our method respectively, and compared the precision and resolution of the target features in the images to demonstrate the advantages of using time-frequency images to process the low-altitude UAV radar echo data. On the other hand, heat maps are utilized to represent the output feature representation vector obtained by PCA, CNN, and the feature extraction network based on self-encoder and contrast learning proposed in this paper.

For radar UAV detection performance, detection rate, false alarm rate and missing alarm rate are three common metrics for accuracy evaluation. Supposed that among all the samples to be identified, the number of samples that do not contain the UAV target is N1, the number of samples for UAV is N2, the number of samples detected as UAV target is and the number of samples that do not contain the target is. Then the metrics can be expressed as:

$$p_d = \frac{|x_i|(L(x_i)=0, p(x_i)=0) \cup (L(x_i)=1, p(x_i)=1) \cup (L(x_i)=2, p(x_i)=2)|}{N_1 + N_2}$$

$$p_f = \frac{|x_i|(L(x_i)=0, p(x_i)=1) \cup (L(x_i)=0, p(x_i)=2) \cup (L(x_i)=1, p(x_i)=2)|}{N_2'}$$

$$p_m = \frac{|x_i|(L(x_i)=1, p(x_i)=0) \cup (L(x_i)=2, p(x_i)=0)|}{N_2} \quad (17)$$

, where $p_d$, $p_f$, and $p_m$ denotes detection rate, false alarm rate, and missing alarm rate [39] respectively. $x_i$ indicates any sample being tested, and $p(x_i)$ and $L(x_i)$ are the predicted and true label of $x_i$. p=0 if no UAV target is detected, p=1 if one UAV target is detected, p=2 if two UAV targets are detected. It can be seen that the smaller $p_f$, $p_m$ is, the more accurate the UAV is discovered, and the higher $p_d$ is, the more accurate the UAV is detected.

### 4.4. Experiments on Feature Extraction

Select a section of simulation data with a strong ground clutter background (SNR=-3dB): Initial distance between radar and UAV target is 800m, and initial UAV altitude is 50m with a radial velocity of 20m/s. Besides the simulation data, a section of real measured data with a strong ground clutter background (SNR=3dB) is also selected: The initial distance between the radar and the UAV-1 with a speed 17m/s is 0.15Km, and the distance between the radar and the UAV-2 with a speed 24m/s is 1.22Km. Fig. 8 shows the RD map of radar echo mentioned above for CA-CFAR target detection after pulse phase reference accumulation with a slow time step window length of 512.

Based on the location cell of the searched peak in the distance-velocity two-dimensional plane, it was estimated that the initial distance to the target was 797.7m and the radial velocity of the target was 19.6m/s in the simulation data segment. However, due to the signal fading caused by the dramatic change in the target echo amplitude, the accumulated peak at this point is 0.2669e4, and the accumulation results in a blind velocity partials peak of 0.2314e4 close to the main peak, resulting in an increase in the false alarm rate of the UAV detection and reducing the accumulation gain. RD plots based on real measured data also suffer from the same main flap spreading problem described above.

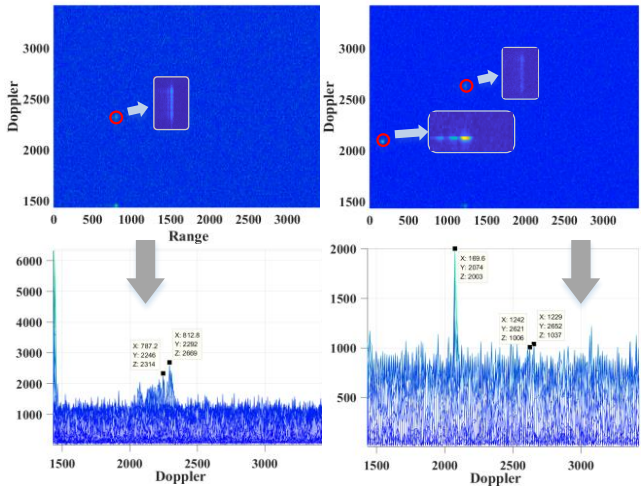

*Fig. 8. RD map of real simulated and real measured radar echo signal*

The time-frequency images with corresponding reconstructed binary diagrams obtained using the ZAM transform and the Short Time Fourier Transform (STFT) method respectively for the simulated signals in Fig. 8 are shown in Fig. 9. The radar carrier frequency of the real measured data is 35 GHz and the initial Doppler frequency is 2433.3Hz obtained by quantification of the UAV velocity, which is generally consistent with the frequency line 2470 Hz in the time-frequency diagram in Fig. 11. From the time-frequency plot, it can be seen that the ZAM transform has a much better resolution than the STFT. The binary plot of the self-encoder reconstruction shows that the data reconstructed by ZAM has a higher target resolution and less ground clutter than STFT.

To further demonstrate the performance benefits of introducing contrast learning methods for data dimensionality reduction and feature extraction of time-frequency images in the background of strong clutter, heat maps [40] are utilized to reflect the similarity between the feature vectors extracted from CNN networks in [37], PCA in [35], and the module proposed in this paper.

The vertical axis of the heat map represents the number of sample signals, and the horizontal axis represents the length of the feature point, where the color represents the amplitude intensity of each cell, the brighter the color, the higher the amplitude. The experiment takes 45 signals from different categories with labels 0, 1, and 2, each with 15 signals. The first 20 bits of the feature vector are the output of the encoder based on contrast learning and the last 30 bits are the output of the encoder based on image reconstruction. It can be seen from Fig. 10(b) that the magnitude of the feature vectors of the echo signal with contrastive learning containing the UAV target (label= 1 or 2) has a stronger positional correlation than that from CNN and PCA, indicating that the signals differ very little in terms of the features reflecting the number of targets, while the distribution of the feature vector amplitudes of signals with and without targets appear uncorrelated.

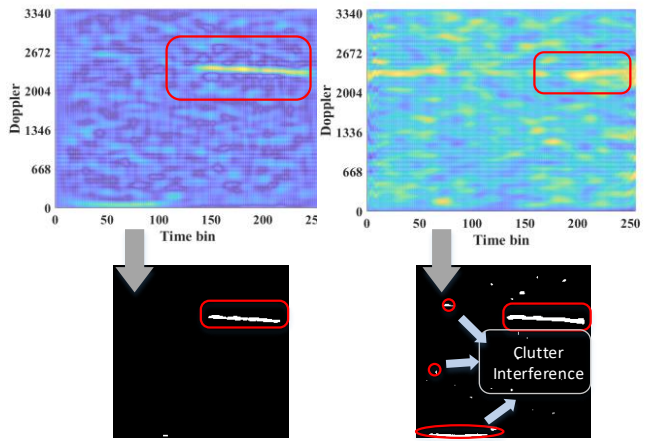

*Fig. 9. Time-frequency images with corresponding reconstructed binary diagrams obtained using the ZAM transform and the Short Time Fourier Transform (STFT) method*

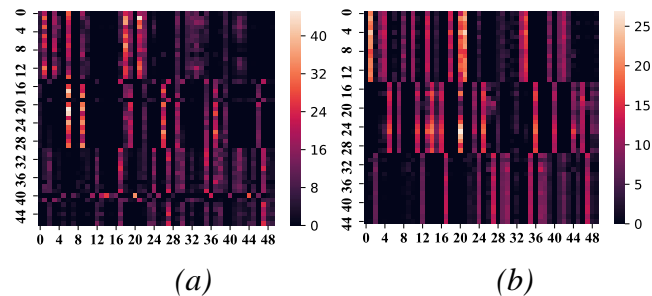

*(a)                    (b)*

*Fig. 10. Feature heat map of the same sample signal*

### 4.5. Experiments on UAV detection
### 1. Experiments on Simulated Data

In the two scenarios of strong ground clutter and weak interference, 2000 groups of UAV radar echo data of each SNR with an observation time of 500ms, and the amount of data from no target, single target, and the double target is balanced in each scenario. The results of the comparison between the experimental and comparison methods with different signal-to-noise ratios when   are shown in Fig.11(a) and Fig.11(b).

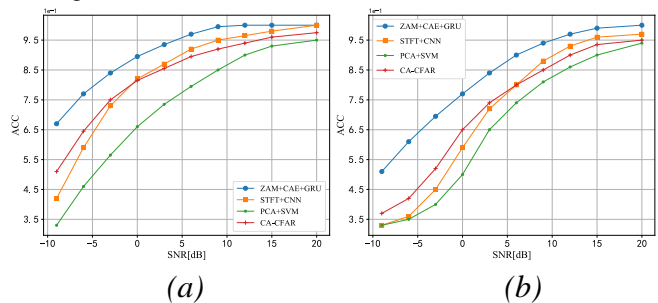

*(a)                    (b)*

*Fig. 11. Detection Accuracy of the comparison between the experimental and comparison methods*

The performance of the experimental method was better than that of the comparison methods in the weak ground clutter interference scenario. However, the difference in performance between the two groups is not large (about 6% for SNR greater than -3dB), and the difference gradually decreases as the SNR increases; in addition, the overall recognition rate of the target echo in the strong ground clutter interference scenario is still about 85% for the experimental group even in the SNR= -3dB environment, while the performance of the control group decreases faster as the SNR increases. This shows that the proposed method can effectively identify low-altitude UAV targets under strong clutter background.

In Table 3, different thresholds are used for the likelihood of belonging to a drone in the classification output to obtain data on the detection and miss rates of each model at different false alarm rates when SNR=-3dB.

**Table 3** Classification outcomes

| Method \ DA | $p_f$ | | | |
|---|---|---|---|---|
| | $10^{-3}$ | $10^{-2}$ | $10^{-1}$ | 0.4 |
| ZAM+CAE+GRU | **0.7681** | **0.8625** | **0.9280** | **0.9880** |
| CFAR+RNN | 0.6533 | 0.8119 | 0.8736 | 0.9536 |
| CWD+SAE+SVM | 0.5943 | 0.7401 | 0.8260 | 0.9260 |
| PCA+SVM | 0.5006 | 0.7875 | 0.8513 | 0.9213 |

As seen from the Table 4, the detection rate of the control and experimental models increases with the false alarm rate when the signal-to-noise ratio is low, with the method in this paper maintaining the lowest false alarm rate at detection rates above 85%.

**Table 4** Detection and miss rates of each model at different false alarm rates when SNR=3dB

| Method | $p_f = 10^{-2}$ | | $p_f = 10^{-1}$ | | $p_f = 0.4$ | |
|---|---|---|---|---|---|---|
| | $p_d$ | $p_m$ | $p_d$ | $p_m$ | $p_d$ | $p_m$ |
| CA-CFAR | 0.5073 | 0.4529 | 0.8332 | 0.1389 | 0.9018 | 0.0910 |
| PCA+SVM | 0.4551 | 0.4018 | 0.7975 | 0.2022 | 0.8276 | 0.1721 |
| STFT+CNN | 0.3926 | 0.3896 | 0.8318 | 0.1550 | 0.8831 | 0.1005 |
| ZAM+CAE+GRU | **0.6790** | **0.2118** | **0.9042** | **0.0402** | **0.9404** | **0.0212** |

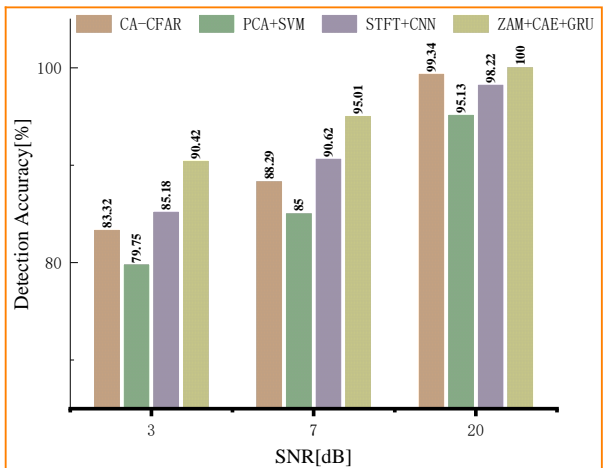

*Fig. 12. Detection Accuracy on real measure data*

### 2. Experiments on Real Data

Fig. 12 shows the detection results of the four methods on real data with $p_f = 10^{-1}$. Specifically, [35], [36], and [37] can only achieve a $p_d$ of 79%-85% when SNR= 3dB. Meanwhile, in Table IV, they have $p_m$ up to 45.2%. In contrast, the $p_m$ of the proposed method declines from 21.18% to 2.12%, while the $p_d$ increases from 67.90% to 94.04%. In a word, the comparison methods usually produce high $p_m$ in strong clutter, and the false detection rate increases significantly with a higher detection rate. By introducing a contrastive learning module and ZAM transform, the proposed method can reach a balance between $p_d$, $p_m$ and $p_f$.

## 5. Conclusions

In this paper, an UAVs detection system based on deep contrast learning is proposed to solve the problem of reliable identification of small UAVs at low-altitude in the environment of strong ground clutter: A low-altitude UAV radar echo model under multipath interference is first established. Based on the echo components and the UAV Doppler domain identifiable mechanism, a time-frequency transform method using ZAM transform and image enhancement is proposed. In addition, to improve the accuracy and precision of the detection, a feature extraction module based on contrast learning was used to eliminate ground clutter and time-frequency feature fusion, and a detection module was designed based on the corresponding features, while the performance of the model was verified by analyzing both the real and simulated datasets. Finally, the experimental results show that, under different signal-to-noise ratios and different intensity of ground clutter interference, the system has a lower false alarm rate than the comparison algorithm and higher reliability of signal feature extraction and accuracy of UAV target recognition.